\documentclass[aps,pra,twocolumn,showpacs,groupedaddress,letterpaper]{revtex4-1}

\usepackage{graphicx}
\usepackage{subfigure}
\usepackage{color}
\usepackage{physymb}
\usepackage{enumitem}
\usepackage{mdframed}
\usepackage{framed}
\usepackage{epstopdf}

\begin{document}

\title{Upper-division student difficulties with Separation of Variables}


\author{Bethany R. Wilcox}
\affiliation{Department of Physics, University of Colorado, 390 UCB, Boulder, CO 80309}

\author{Steven J. Pollock}
\affiliation{Department of Physics, University of Colorado, 390 UCB, Boulder, CO 80309}

\begin{abstract}
Separation of variables can be a powerful technique for solving many of the partial differential equations that arise in physics contexts.  Upper-division physics students encounter this technique in multiple topical areas including electrostatics and quantum mechanics.  To better understand the difficulties students encounter when utilizing the separation of variables technique, we examined students' responses to midterm exam questions and a standardized conceptual assessment, and conducted think-aloud, problem-solving interviews.  Our analysis was guided by an analytical framework that focuses on how students activate, construct, execute, and reflect on the separation of variables technique when solving physics problems.  Here we focus on student difficulties with separation of variables as a technique to solve Laplace's equation in both Cartesian and spherical coordinates in the context of junior-level electrostatics.  Challenges include: recognizing when separation of variables is the appropriate tool; recalling/justifying the separated form of the potential and the need for the infinite sum; identifying implicit boundary conditions; and spontaneously reflecting on their solutions.  Moreover, the type and frequency of errors was often different for SoV problems in Cartesian and spherical geometries.  We also briefly discuss implication of these our findings for instruction.  
\end{abstract}

\maketitle

\section{Introduction}

Research into student difficulties at the upper-division level is a growing area of Physics Education Research \cite{meltzer2012resource}.  Students in upper-division courses are asked to manipulate increasingly sophisticated mathematical tools as they tackle more advanced physics content.  Because of this, some of the literature around student difficulties at this level has focused on the use of mathematical tools and techniques during physics problem-solving \cite{caballero2014mathphys}.  For example, one mathematical technique that appears repeatedly throughout the undergraduate physics curriculum is separation of variables (SoV) as a method for solving Partial Differential Equations (PDEs).  

PDEs appear in multiple contexts in the upper-division, undergraduate physics curriculum (e.g., waves on a string, Maxwell's equations, the Schr\"{o}dinger equation).  One of the most common approaches to solving PDEs in physics is to turn them into multiple Ordinary Differential Equations (ODEs) using a technique known as separation of variables (SoV).  Here, we use the term SoV to refer to the technique of guessing a general solution with a functional form that allows the PDE to be separated into several ODEs, and then solving these ODEs individually with appropriate boundary conditions.  This technique is not to be confused with the strategy, also conventionally referred to as separation of variables, used to solve separable ODEs by isolating terms with the function on one side of the equals sign and the independent variable on the other side and integrating both. 

Some existing work has been done investigating student difficulties around solving ODEs in both physics and mathematics.  Much of this differential equations literature focuses on students' use of graphical techniques to solve linear ODEs \cite{rasmussen2001de, rasmussen2000de, habre2000de}, or numerical solutions to more complex differential equations that cannot be solved analytically \cite{duda2011pbl}.  However, we are not aware of any existing research specifically targeting student difficulties with the SoV technique.  

At the University of Colorado Boulder (CU), physics students encounter SoV several times in their undergraduate courses.  The first exposure is often in sophomore classical mechanics as a technique to solve Laplace's equation in Cartesian coordinates, typically in the context of finding the temperature as a function of position in a mechanical system with given boundary conditions.  Students may also encounter this technique in a differential equations course taken from the Math department.  Junior electrostatics is the next common place where CU's physics majors see SoV in the context of solving Laplace's equation for the electric potential, $V$, in 2D and 3D Cartesian coordinates and spherical coordinates with azimuthal symmetry.  Students do not typically encounter spherical SoV with $\phi$-dependent solutions until quantum mechanics where it is used to solve the Schr\"{o}dinger equation for the hydrogen atom.  In discussions with the physics faculty at CU, some instructors have expressed concern that students do not begin to demonstrate mastery of the SoV technique until they see it for a third time in quantum mechanics, and sometimes not even then.  

In this paper, we focus on students' use of SoV in junior-level electrostatics as a technique to solve Laplace's equation in both Cartesian coordinates (i.e., $\nabla^2V(x,y,z)=0$) and spherical coordinates with azimuthal symmetry (i.e., $\nabla^2V(r,\theta)=0$).  These types of problems typically ask for an expression for the voltage in a charge-free region and provide an expression for the voltage along the boundary of that region.  Given its many possible uses, we do not claim that the research presented here will span the space of all possible difficulties with SoV, but rather will provide a sampling of the types of challenges students encounter when dealing with the SoV technique in electrostatics.  

In this paper we utilize an analytical framework \cite{wilcox2013acer} describing the use of mathematical tools in physics problem solving to structure our investigation and analysis of student difficulties with SoV (Sec.\ \ref{sec:methods}).  We then present our findings, including common difficulties we identified in our student population and a brief discussion of implications for instruction (Sec.\ \ref{sec:results}).  We end with limitations and future work (Sec.\ \ref{sec:discussion}).

\section{\label{sec:methods}Methods}

Problem solving at the upper-division level is often long and complex, and making sense of students' work around these upper-division problems can be difficult.  There is a large variety of potential moves and/or errors that students can make at different stages of a problem, and these moves can impact the remainder of their solutions in unpredictable ways.  To help manage this complexity, we make use of an analytical framework known as ACER (Activation, Construction, Execution, Reflection) to scaffold our analysis of student difficulties with SoV \cite{wilcox2013acer}.

\subsection{\label{sec:acer}The ACER Framework}

The ACER framework organizes the problem-solving process into four general components: \emph{activation} of mathematical tools, \emph{construction} of mathematical models, \emph{execution} of the mathematics, and \emph{reflection} on the results. These components were identified by studying expert problem solving \cite{wilcox2013acer} and are grounded in both a resources \cite{hammer2000resources} and epistemic framing \cite{bing2008thesis} perspective on the nature of knowledge.  However, while the general structure of experts' back-of-the-book style problem solving may be reasonably context independent, the specific details of how a particular mathematical tool is used in upper-division problem solving is often highly dependent on the context in which that tool is being used.  For this reason, the ACER framework was designed to be operationalized for specific mathematical tools in specific physics contexts.  The operationalization process results in a researcher-guided outline of key elements in a correct and complete solution to a particular problem or set of problems.  The process of operationalizing ACER for SoV will be discussed in greater detail in Sec.\ \ref{sec:Sacer}, and additional details about the ACER framework can be found in Ref.\ \cite{wilcox2013acer}.

\subsection{\label{sec:context}Study Context}

Data for this study were largely collected from the first half of a two-semester Electricity and Magnetism sequence
at the University of Colorado Boulder (CU). This course, called E\&M 1, typically covers the first 6 chapters of Griffiths' text \cite{griffiths1999em} (i.e., electrostatics and magnetostatics).  Students in this course are junior and senior-level physics, astrophysics, and engineering physics majors with a typical class size of 30-70 students. At CU, E\&M 1 is often taught with varying degrees of interactivity through the use of research-based teaching practices including peer instruction using clickers \cite{mazur1997pi} and tutorials \cite{chasteen2012transforming}.  We collected data from three distinct sources for this investigation: student solutions to instructor designed questions on traditional midterm exams, responses to two question from the multiple-response Colorado Upper-division Electrostatics (CUE) Diagnostic \cite{wilcox2014cmr}, and think-aloud, problem-solving interviews.  Ultimately, exams provided quantitative data identifying common difficulties and interviews offered deeper insight into the nature of those difficulties

Midterm exam data were collected from 10 semesters of the E\&M 1 course (N=474) taught by 8 different instructors.  Of these, six were traditional research faculty and two were physics education researchers.  Two of these instructors, including one of the PER faculty members (SJP), taught the course twice during data collection.  Questions on the exams were developed solely by the instructor for that semester.  In four cases, the instructor asked one or more SoV questions on both the midterm and final exams, thus the following section reports the analysis of 15 distinct exam questions for a total of N=744 unique solutions.  As our goal is to identify the presence of common student difficulties, the remainder of this analysis will report N as the number of solutions rather than number of students.  

\begin{figure}
\begin{minipage}{.95\linewidth}
\flushleft \emph{A rectangular pipe, running parallel to the z-axis, extending from $-\infty$ to $\infty$, has three grounded metal sides and a fourth side maintained at a constant potential, $V_o$, as in the figure.  \\  Find the potential $V(x,y)$ at all points inside the pipe.  }
\begin{center}
\includegraphics[width=1.5in, height=1in]{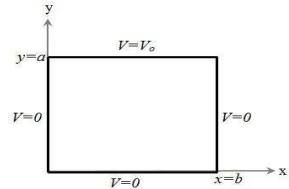}
\end{center}
\end{minipage}
\caption{An example of a canonical exam problem targeting Cartesian SoV.  Variations on this question in our data include providing a non-constant potential on the forth side [e.g., $V_o(x,y=a)=V_o sin\hspace{1mm}\pi x/a$], placing the non-grounded side at $y=0$ rather than $y=a$, or placing one of the grounded sides at infinity.  } \label{fig:examqSoVc}
\end{figure}

\begin{figure}
\begin{minipage}{.6\linewidth}
\flushleft \emph{A thin spherical shell of radius $R$ and centered on the origin has a voltage on its surface given by $V(R,\theta)=V_o cos^2(\theta)$.  \\  \vspace{2mm}Find the potential $V(r,\theta)$ everywhere (inside \emph{and} outside the sphere).}
\end{minipage}
\begin{minipage}{.35\linewidth}
\includegraphics[width=1in, height=.9in]{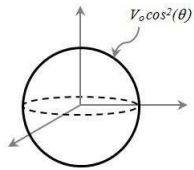}
\end{minipage}
\caption{An example of a canonical exam problem targeting spherical SoV.  Variations on this question include providing a simpler or more complex expression for the boundary (e.g., $V(R,\theta)=V_o cos\hspace{1mm}\theta$), giving the boundary condition in terms of Legendre polynomials (e.g., $V(R,\theta)=P_9(cos\hspace{1mm} \theta$)), or only asking for the potential inside or outside the sphere.  } \label{fig:examqSoVs}
\end{figure}

Exam questions requiring the use of SoV in Cartesian and spherical coordinates are both common at CU.  Four of the exam questions in our sample (N=235 solutions from 3 semesters) provided students with a rectangular pipe or gutter with given values for the voltage on each side and asked for an expression for the voltage valid everywhere inside (e.g., Fig.\ \ref{fig:examqSoVc}).  The remaining 11 exam questions (N=509 solutions from 9 semesters) provide students with an azimuthally symmetric expression for the voltage on the surface of a spherical shell and asked for an expression for the voltage valid inside and/or outside the shell (e.g., Fig.\ \ref{fig:examqSoVs}).  

\begin{figure}[b]
\begin{minipage}{.6\linewidth}
\flushleft \emph{{\bf Q13} - To solve for V inside the box by separation of variables, which form of the solution should you choose?  \\ \vspace{3mm}Select only one.}
\end{minipage}
\begin{minipage}{0.35\linewidth}    
      \begin{center}
        \includegraphics[width=25mm, height=22mm]{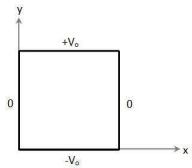}
      \end{center}
\end{minipage}
\begin{minipage}{.96\linewidth}
      \flushleft
      \begin{enumerate}[label=\Alph*., leftmargin=6mm, itemsep=-1mm]
      \item $V(x,y) = (A \hspace{1mm}e^{ky}+ B \hspace{1mm}e^{-ky})\dot(C \hspace{1mm}sin\hspace{1mm}kx + D \hspace{1mm}cos \hspace{1mm}kx)$
      \item $V(x,y) = (A \hspace{1mm}sin \hspace{1mm}ky + B \hspace{1mm}cos \hspace{1mm}ky)\dot(C \hspace{1mm}e^{kx}+ D \hspace{1mm}e^{-kx})$
      \item $V(x,y) = (A \hspace{1mm}sin \hspace{1mm}ky + B \hspace{1mm}cos \hspace{1mm}ky)\dot(C \hspace{1mm}sin \hspace{1mm}kx + D \hspace{1mm}cos \hspace{1mm}kx)$
      \item $V(x,y) = (A \hspace{1mm}e^{ky}+ B \hspace{1mm}e^{-ky})\dot(C \hspace{1mm}e^{kx}+ D \hspace{1mm}e^{-kx})$
      \item \emph{More than one of these could be used}
      \end{enumerate}
\end{minipage}
\caption{The multiple-response CUE question related to the Cartesian SoV.  The prompt has been shortened and paraphrased, see Ref.\ \cite{wilcox2014cmr} for full prompt.}  \label{fig:cmrqSoVc}
\end{figure}

Student responses to the multiple-response CUE provide an additional data source.  Two questions on the CUE deal with SoV: one in Cartesian (Fig.\ \ref{fig:cmrqSoVc}), and one in spherical.  The spherical CUE question was a multiple-response version of a prompt like that in Fig.\ \ref{fig:examqSoVs}, but, rather than solving for the potential, students were asked only to select the appropriate solution method and to justify their choice (see Ref.\ \cite{wilcox2014cmr} for exact prompt).  CUE data were collected from four semesters (N=145) of the E\&M 1 course at CU; three of these were courses for which we also have exam data.  In addition to the CU data, we also collected multiple-response CUE data from 9 courses at 7 external institutions (N=161) ranging from small liberal arts colleges to large research institutions.  

Think-aloud interviews (N=11) were conducted in two sets performed roughly two years apart in order to further probe preliminary difficulties identified in student exams.  The first interview set (N=4) was designed to target student difficulties with SoV in Cartesian coordinates and was conducted prior to the development of the ACER framework.  The students were asked to determine the voltage inside a semi-infinite plate with one side held at a constant potential (similar to the question in Fig.\ \ref{fig:examqSoVc}).  The students were directly prompted to approach this problem by using the SoV technique to solve Laplace's equation.  They were also provided the expression for $\nabla^2$ in Cartesian coordinates along with the solutions to the relevant ODEs and the integral expression needed to determine the coefficients in a Fourier series.  From the perspective of the ACER framework, this prompt clearly targeted the Construction and Execution components, but bypassed Activation.  

The second interview set (N=6) began with a spherical SoV problem in which the students were given a spherical shell with a known voltage on the surface (see Fig.\ \ref{fig:examqSoVs}).  To directly target Activation, the prompt did not specifically mention Laplace's equation or prompt the students to use the SoV technique.  Students who were able to complete this problem in the allotted time (N=3) were also explicitly prompted to come up with a way to check their solution in order to convince themselves their expression was correct (i.e., Reflection).  The second interview set ended by asking students to begin working though the Cartesian question shown in Fig.\ \ref{fig:examqSoVc}, though no student had time to fully complete this question. While the Cartesian prompt also did not directly prompt SoV, this question provided minimal insight into spontaneous Activation because it came immediately after a spherical SoV question.

\subsection{\label{sec:Sacer}Operationalizing ACER}

The process of operationalizing ACER is presented in detail in Ref.\ \cite{wilcox2013acer}.  Briefly, in order to operationalize the framework, a content expert utilizes a modified form of task analysis \cite{catrambone2011taps, wilcox2013acer} in which they work through the problems of interest while carefully documenting their steps and mapping these steps onto the general components of the framework.  Additional content experts then review and refine the resulting outline until consensus is reached that the key elements of the problem have been accounted for.  This expert-guided scheme then serves as a preliminary coding structure for analysis of student work.  If necessary, the operationalization can be further refined to accommodate aspects of student problem-solving that were not captured by the expert task analysis.  

To guide our data collection and analysis, we operationalized the ACER framework for problems like those in Fig.\ \ref{fig:examqSoVc} and Fig.\ \ref{fig:examqSoVs}.  The elements of the operationalized ACER framework are detailed below. Element codes are for labeling purposes only and are not meant to suggest a particular order, nor are all elements always necessary for every problem.  In particular, the elements of Construction and Execution are unlikely to occur in the specific order listed as experts can, and often do, iterate back and forth between deriving and solving ODE's and identifying and matching boundary conditions.  

\emph{Activation of the tool --}  The first component of the framework involves identifying SoV as the appropriate mathematical technique to solve for the voltage.  We identified three elements in the form of cues present in a prompt that are likely to activate resources associated with SoV.  

\vspace{-2mm}
\begin{enumerate}[label={\bf A\arabic*}:, align=left] \itemsep0pt \parskip0pt \parsep0pt
  \item The question provides boundary conditions and asks for separation of variables directly or provides the expression for the general solution
  \item The question provides boundary conditions and uses language associated with separation of variables (e.g., infinite sum, Legendre polynomials, Fourier series, general solution)
  \item The question provides boundary conditions and asks for the electric potential or voltage in a charge free region
\end{enumerate}
\vspace{-2mm}

For element A1, it is more common for a question to provide the general solution for the voltage in spherical geometries than in Cartesian, in part because the functional form of the general solution in Cartesian depends on the boundary conditions (element C2).  

\emph{Construction of the model --}  Elements in this component deal with modifying the general expression for the solution to Laplace's equation so that it matches the boundary conditions.  As there is no ambiguity in the signs of the separation constants for SoV in spherical coordinates due to the nature of the $\theta$ coordinate, the second element of Construction is specific to Cartesian problems. 

\vspace{-2mm}
\begin{enumerate}[label={\bf C\arabic*}:, align=left] \itemsep0pt \parskip0pt \parsep0pt
  \item Express all relevant boundary conditions, both those explicitly given in the prompt/figure and those implicit from the physical situation (e.g., $V(r\rightarrow \infty) \rightarrow 0$)
  \item (Cartesian only) Choose the signs of the separation constants or select the functional forms for the solution that are appropriate for the boundary conditions
  \item Apply each boundary condition to the general solutions in order to solve for all unknown constants (set up only)
\end{enumerate}
\vspace{-2mm}

Note that these elements do not necessarily occur sequentially, either with respect to one another or with respect to elements of the Execution component.  

\emph{Execution of the mathematics --}  This component of the framework deals with elements involved in executing the mathematical operations related to SoV.   As students are rarely (if ever) asked to actually produce and solve the ODEs resulting from SoV in spherical coordinates, the first two elements of Execution are specific to Cartesian (and potentially Cylindrical) problems.  

\vspace{-2mm}
\begin{enumerate}[label={\bf E\arabic*}:, align=left] \itemsep0pt \parskip0pt \parsep0pt
  \item (Cartesian only) Manipulate the PDE into ODEs using the separated form of the potential (e.g., $V(x,y,z)=X(x)Y(y)Z(z)$)
  \item (Cartesian only) Know or look up the solution to these ODEs given the signs of the separation constants
  \item Calculate values for all unknown constants based on applying the boundary conditions through zero matching, term matching, or `Fourier's trick' integrals
  \item Manipulate algebraic expressions and compile an interpretable expression for $V(\vec{r})$
\end{enumerate}
\vspace{-2mm}

Element E3 can be accomplished using a variety of strategies sometimes involving several smaller steps depending on the particular boundary conditions.  These strategies, which we refer to as zero matching, term matching, and Fourier's trick, are explicitly described in Sec.\ \ref{sec:resultsSoVc}.

\emph{Reflection on the result --} The final component includes elements related to checking and interpreting aspects of the solution, including intermediate steps and the final result.  While many different techniques can be used to reflect on a physics problem, the following four are particularly common when dealing with SoV.  
\vspace{-2mm}
\begin{enumerate}[label={\bf R\arabic*:}, align=left] \itemsep0pt \parskip0pt \parsep0pt
  \item Check the units of the final expression
  \item Check that the solution matches all boundary conditions
  \item (Spherical only) Check that the solution behaves as expected in known limits
  \item Confirm that the solution satisfies Laplace's Equation
\end{enumerate}
\vspace{-2mm}

Element R3 refers specifically to checking the functional dependence, rather than the value, of the voltage in known limits.  For example, checking that $V(r\rightarrow \infty)=0$ would be considered R2 while showing that $V$ goes to zero as $1/r$ would be R3.  The final element of Reflection (R4) was added to the framework after initial analysis of student work where we observed that mistakes in the Construction and Execution components sometimes resulted in solutions that did not satisfy Laplace's equation.  

In the next section, we will apply this operationalization of ACER to investigate student work on canonical SoV problems in electrostatics.  

\section{\label{sec:results}Findings}

While we intentionally operationalized the framework for problems in both Cartesian and spherical geometries, our analysis found that the type and frequency of errors was often quantitatively and/or qualitatively different for the two geometries.  For this reason, we report on student difficulties with these two geometries separately.  Where appropriate, we also synthesize these findings in Sec.\ \ref{sec:discussion}.  

\subsection{\label{sec:resultsSoVc}\emph{Student difficulties with Separation of Variables in Cartesian Coordinates}}

This section presents the identification and analysis of common student difficulties with the separation of variables technique in Cartesian coordinates organized by component and element of the operationalized ACER framework (Sec.\ \ref{sec:Sacer}).  

\subsubsection{\label{sec:activation}Activation of the tool}

Canonical Cartesian SoV questions (e.g., Fig.\ \ref{fig:examqSoVc}) are highly distinctive both because they provide boundary conditions and because they do not provide information needed to solve the problem using other common, analytical methods (e.g., $\rho(\vec{r})$).  Elements A1-A3 describe different types of prompts that can cue students to activate resources related to SoV, loosely organized according to the likelihood that they will do so.  One of the four exam questions (N=69) explicitly prompted students to use SoV and presented them with the appropriate separated form for the solution (i.e., $V(x,y)=X(x)Y(y)$); thus, these solutions provide minimal insight into Activation.  Alternatively, prompts consistent with A2 or A3 require students' to identify SoV as the appropriate technique, and only one of the 166 solutions to implicit prompts used a method other than SoV.  Due to the distinctive nature of Cartesian SoV questions, we are hesitant to interpret this result as evidence that our students have a solid understanding of \emph{when} SoV is the correct approach; however, it does suggest they consistently activate resources related to SoV in response to these canonical questions.  

\subsubsection{\label{sec:constructionc}Construction of the model}

The Construction component deals with mapping between the physics and mathematics of a problem.  For SoV, this process includes identifying all necessary boundary conditions (element C1).  Cartesian SoV problems typically provide these boundary conditions explicitly in the prompt (see Sec.\ \ref{sec:resultsSoVs} for discussion of implicit boundary conditions).  Consistent with this, only a small fraction of solutions (5\%, N=12 of 234) used an incorrect value or expression for the boundary conditions.  Common errors included putting the non-zero boundary condition on the wrong side (N=4), including inappropriate implicit boundary conditions (N=4, e.g., $V(x\rightarrow \infty)=0$), or listing the value of the potential at a point (usually a corner) rather than along a side (N=2).  One interview participant also listed the boundary conditions at each corner.  This student recognized and corrected the error after attempting to apply a boundary condition and finding that this did not help him solve for any unknowns.  Thus, extracting boundary conditions from the prompt or figure for Cartesian SoV questions was not a significant stumbling block for the majority of our students.  

After identifying boundary conditions, the next step is to produce a general expression for the voltage that can satisfy these boundaries (element C2).  For Cartesian SoV questions, this amounts to deciding which direction ($x$ or $y$) gets the exponential dependence.  One exam prompt (N=69) asked students to select the appropriate general solution for $V(x,y)$ from the two possible Cartesian solutions to Laplace's equation, and all students selected the correct expression.  The multiple-response CUE asked a similar question but provided two additional response options which featured sinusoidal or exponential dependence in both directions (Fig.\ \ref{fig:cmrqSoVc}).  In contrast to the results on the exam question, only two-thirds of CU students (66\%, N=95 of 145) selected the correct expression, while the majority of the remaining students selected either the solution with flipped functional dependence (12\%, N=17 of 145) or one of the two response options that did not satisfy Laplace's equation (14\%, N=21 of 145).  This trend is even more pronounced in student populations at other institutions, with almost a quarter of students (23\%, N=37 of 161) selecting either purely sinusoidal or purely exponential dependence.  

The remaining three exam prompts did not provide possible expressions for the voltage.  In practice, this meant that students could explicitly work through the process of separating Laplace's equation (elements E1, C2, and E2) or jump straight to a general expression for the potential without deriving this expression.  Using the former strategy, element C2 requires deciding which separation constant gets the negative value (and thus which direction gets the sinusoidal dependence).  Roughly two-thirds of solutions (71\%, N=117 of 165) explicitly commented on the signs of the separation constants, and most (87\%, N=102 of 117) assigned the negative constant such that it was consistent with the boundary conditions.  Similarly, just over three-quarters of solutions that jumped straight to a general expression (79\%, N=38 of 48) also gave a functional form that was consistent with the boundary conditions.  As with the responses on the multiple-response CUE, the most common errors included either flipping the functional form or having sinusoidal solutions (or negative separation constants) in both directions.  

Of the six interview students who progressed far enough in their solution to begin one of the Cartesian SoV questions, four derived the general expression directly from Laplace's equation.  All four correctly identified which coordinate ($x$ or $y$) should be given the negative sign and justified this based on the boundary conditions.  The two remaining students jumped straight to a general expression for the potential without derivation.  One of these two argued for needing both exponential and sinusoidal dependence, but used complex rather than real exponentials.  The second of these students argued for sinusoidal behavior in both $x$ and $y$ directions and justified this by stating that the boundary conditions in both directions could be matched by sines.  Only when directly asked to show that this expression solved Laplace's equation did this student recognize that one of the two directions must have exponential dependence.  This result suggests that students who argued for purely sinusoidal or purely exponential dependence on the exams and CUE may have been focusing on satisfying the boundary conditions while failing to consider that, ultimately, the solution must also satisfy Laplace's equation.  

The final element of Construction deals with setting up the equations that are used to solve for the unknown constants in the general solution in order to match the boundary conditions (note, algebraic mistakes related to solving these equations will be discussed in relation to the Execution component).  Almost a third of solutions (30\%, N=71 of 234) included issues with setting up the equations to solve for one or more constants.  The exact details of these errors were often strongly tied to the nature of the specific boundary conditions in question, but common issues included (see Table \ref{tab:Csc3}):  inappropriately arguing that one exponential term should be eliminated, incorrectly setting up or failing to utilize the non-zero boundary condition, or setting up an integral (i.e., Fourier's trick) incorrectly.  As we might expect, the majority of these issues centered around setting up the expression to match the non-zero boundary condition.  

\begin{table}
\caption{Difficulties setting up expressions match the boundary conditions (BCs) in Cartesian. To account for the fact that certain difficulties were not applicable to all exam prompts, percentages are given with respect to the subset of incorrect solutions taken from the applicable semesters.  Codes are not exhaustive or exclusive but represent the most common themes, thus the total N in the table need not sum to 71.}\label{tab:Csc3}
  \begin{ruledtabular}
    \begin{tabular} {p{6cm} c c}
	\textbf{Difficulty} 				& \textbf{N} 	& \textbf{Percent} 					\\
	\hline
	 Incorrect application of non-zero BC 	& 32 		& 20\% 							\\
           \hspace{3mm} e.g., not plugging in a value for $x$ or     & 	& (of N=71)						\\
           \hspace{10mm} $y$, or  using $Y(y=a)=V(x)$    & 	&								\\
           Missing + or - exponential term	 	& 16 			& 24\% 						\\
	\hspace{3mm} (box questions only; excludes gutters)		 & 	& (of N= 66)				\\
	Never applied non-zero BC  	& 14 			& 20\% 							\\
	\hspace{3mm}			& 	& 					(of N=71)				\\
	Incorrect setup of Fourier's trick integral 	& 8 			& 12\%					\\
	\hspace{3mm} e.g., missing/extra sum, not multiplying 	& 	& 	(of N=66)				\\
           \hspace{10mm}  by $sin(n'\pi x/a)$   & 	&										\\
    \end{tabular}
  \end{ruledtabular}
\end{table}

In interviews, four students attempted to solve for unknowns (element C3), and three had difficulty doing so.  Consistent with student performance on exams, the most common issue related to using the non-zero boundary condition to solve for the final constant(s).  None of these three students spontaneously included a sum in their solution even after applying the non-zero boundary condition and finding that the resulting equation could not be solved (e.g., $V(x,y=0)=V_o=A sin(\pi x/a)$, where $A$ is a constant).  After being prompted that they actually had an infinite number of solutions (rather than just one), all three students recalled that they needed to introduce a sum of these solutions, but none could clearly justify how this step helped.  One potential explanation for the increased frequency of this issue relative to the exams is that, on exams, students may be recalling an algorithm that includes introducing a sum but have not internalized the motivation for that sum.  If so, it may also be an indication that students are approaching these SoV problems with an epistemic frame that does not require them to justify their steps physically (e.g., Invoking Authority \cite{bing2008thesis}) but rather encourages them to map their solution to these problems onto those of previous problems.  

\subsubsection{\label{sec:executionc}Execution of the mathematics}

The Execution component deals with the procedural aspects of working through the mathematics of a physics problem.  For Cartesian SoV, this can include the process of separating Laplace's equation into ODEs by assuming the separated form of the potential (i.e., $V(x,y)=X(x)Y(y)$, element E1).  Excluding solutions from semesters where the general solution for the potential was provided, more than half of the exam solutions (59\%, N=98 of 165) explicitly included this process and only a small fraction (10\%, N=10 of 98) had difficulties with it.  The most common error (N=7 of 10) was using different constants in the $x$ and $y$ ODEs, resulting in a solution with the correct functional form that does not satisfy Laplace's equation.  

In interviews, six of seven participants commented on or attempted to work through this process of separating Laplace's equation.  Of these, four students spontaneously suggested assuming the separated form of the potential (element E1), though one student noted that he did not understand the motivation for making this assumption. Of the remaining two participants, one clearly articulated that the goal was to separate Laplace's equation into ODEs, but could not recall how to do this on his own.  The other student neither recalled the separated form nor recognized its purpose without being explicitly told.  Additionally, four of the interviewees either did not recognize that the expression $g(y)+f(x)=0$ implies $g(y)=-f(x)=c$ (where $c$ is a constant), or attempted to apply this logic before having fully separated $x$ and $y$ dependent terms (e.g., arguing $X(x)''Y(y)=c$).  Given that there is little (if any) physical motivation for assuming $V(x,y)=X(x)Y(y)$, it becomes particularly important that students understand the mathematical motivation for this move.  However, interviews suggest that, even when students correctly use this assumption in their solution, they may not have a clear sense of the motivation or justification for this assumption. 

The second element in Execution involves solving the ODEs that result from separating Laplace's equation (e.g., $X(x)''=\pm k^2 X(x)$, where $k^2$ is the separation constant).  More than two thirds of solutions (70\%, N=116 of 165) included an expression for one or more ODEs either derived from Laplace's equation or stated without work.  In practice, it is typical for students to simply write down the solutions to these ODEs either by memory or from an equation sheet, and just under a fifth of solutions (16\%, N=19 of 116) provided a general solution that was inconsistent with the ODE they were solving.  Common mistakes included providing a solutions whose functional form was inconsistent with the sign of the separation constant (N=7 of 19), or using the separation constant (rather than its square root) in the expression for the general solution (N=5 of 19).  Thus solving the relatively simple ODEs required for Cartesian SoV questions was not a significant barrier to our students' success.  

The Execution component of ACER also deals with the procedural mathematics of determining values for each of the unknown constants (element E3) in order to match the boundary conditions (element C3).  Our initial analysis of both expert and student work suggested that there were three common strategies used to solve for these constants:

\begin{description}\itemsep0pt \parskip0pt \parsep0pt
\item[Zero matching] Setting unknown constants to zero in order to enforce boundary conditions where $V=0$.
\item[Fourier's trick] The strategy used to solve for the coefficients in a Fourier series by exploiting the integral properties of orthogonal functions.
\item[Term matching] The strategy of exploiting the properties of orthogonal functions to directly match the coefficients of like terms.
\end{description}

Zero matching is nearly always necessary in Cartesian SoV questions, and nearly all of the exam solutions demonstrated some form of zero matching (94\%, N=220 of 234).  Alternatively, whether Fourier's trick or term matching is used to solve for the final unknown constant(s) often depends on the nature of the final boundary condition.  For one of our three exam questions, the final boundary had a constant voltage (e.g., Fig.\ \ref{fig:examqSoVc}), making it necessary to solve using Fourier's trick.  However, the two remaining exams provided a voltage of the form $V(x,y=a)=V_o sin(\pi x/a)$.  In these cases, it is possible to use either Fourier's trick or term matching, though term matching is considerably simpler.  Despite this, more of our students' solutions utilized Fourier's trick (44\%, N=52 of 117) than term matching (36\%, N=42 of 117).  This result may indicate that our students' Cartesian SoV resources are strongly linked to Fourier's trick (rather than term matching), and/or that they have not internalized the properties of orthogonal functions enough to see term matching as a viable strategy.  Without interview data on student reasoning for this specific type of boundary condition, we are not able to distinguish between these two possible explanations.  

When solving for the values of the unknown constants in their general solution (element E3), roughly half of students' solutions (45\%, N=105 of 234) contained various mathematical mistakes.  Common errors included (see Table \ref{tab:Cse3}): losing or gaining a constant factor, incorrectly executing a Fourier's trick integral, (when applicable) not including the constant factor $Y(y=a)$, and not finishing the calculation.  For the two exam questions that could be solved using either Fourier's trick or term matching, the fraction of solutions with mathematical errors was higher in solutions that utilized Fourier's trick (77\%, N=40 of 52) than in those that utilized term matching (33\%, N=14 of 42).  This is likely due, at least in part, to the fact that Fourier's trick requires the set up and execution of an integral, and thus is a more mathematically demanding strategy.

\begin{table}
\caption{Common difficulties when executing the procedural mathematics of solving for constants in the general solution. Percentages are of just the students who exhibited these difficulties (45\%, N=105 of 234).  Codes are not exhaustive or exclusive but represent the most common themes, thus the total N in the table need not sum to 105.}\label{tab:Cse3}
  \begin{ruledtabular}
    \begin{tabular} {p{6.5cm} c c}
	\textbf{Difficulty} 				& \textbf{N} 	& \textbf{Percent} 					\\
	\hline
	 Off by a constant factor or sign		& 33 		&  31\%			\\
           \hspace{3mm} e.g., factor of 2, $V_o$, or length $a$       & 		& 	\\
           Problems with a Fourier's trick integral	 	&  30			& 29\%		\\
	\hspace{3mm} e.g., pulling non-constant terms out of the	 & 	& \\
           \hspace{10mm} integral, or not collapsing the sum & & \\
           Dropping the $Y(y=a)$ factor in the solution	 	&  18			& 17\%		\\
	Not finishing the calculation  	& 27 			& 26\%				\\
    \end{tabular}
  \end{ruledtabular}
\end{table}

For solutions in which the student finished solving for the final constant(s) (78\%, N=130 of 166), it was then necessary to compile all aspects of the solution into a single expression for the voltage (element E4).  Just under a quarter of solutions (22\%, N=29 of 130) either did not compile a final expression or made various mathematical mistakes not related to previous Execution or Construction errors (e.g., incorrectly simplifying exponentials to hyperbolic trig functions, or dropping or adding non-constant factors such as $Y(y)$).  In practice, the interviews provided limited insight into the procedural aspects of solving for the unknown constants as students were, at most, asked to set up the expression for the final constant(s); however, none of our interview participants had difficulty with the simple manipulations required to match the $V=0$ boundaries.  

Ultimately, roughly a quarter of the solutions (26\%, N=61 of 235) included \emph{only} errors related to the elements in the Execution component (i.e., no previous mistakes in Activation or Construction).  This number is high relative to previous research on students' use of mathematical tools where the fraction of students who had difficulty only with Execution was less than a tenth for both multivariable integration (8\% \cite{wilcox2013acer, wilcox2015thesis}) and the Dirac delta function (7\% \cite{wilcox2015delta, wilcox2015thesis}).  Investigations of student difficulties with Taylor Series also found that students made relatively few Execution errors \cite{wilcox2013acer}.  This result suggests that, particularly for problems involving Fourier's trick, the procedural mathematics involved in problems requiring Cartesian SoV can be a significant barrier for our junior-level electrostatics students.

\subsubsection{\label{sec:reflectionc}Reflection on the Result}

The Reflection component deals with the process of checking and/or interpreting the final expression.  It is often the case in Cartesian SoV that mistakes in the Construction or Execution components resulted in an expression for the potential that had the wrong units, did not match the boundary conditions, or did not satisfy Laplace's equation (i.e., elements R1, R2, and R4 respectively).  Overall, we found that very few of our students (N=2 of 234) made explicit, spontaneous attempts reflect on their solution using any of these checks.  This number should be interpreted as a lower bound on the frequency of spontaneous reflections as is possible that more of the exam students made one of these checks and simply did not write it down explicitly on their exam solution.  However, only two of seven interview participants made spontaneous attempts to reflect on their solutions, and exclusively by checking that their general solution satisfied Laplace's equation (element R4).  One other student executed this check only after being directly prompted.  

\begin{table}
\caption{Number of exam students who explicitly utilized each of the three possible reflective checks (N$_{explicit}$) along with the number of solutions that included an error that would have been detected by this check (N$_{incorrect}$).  N$_{total}$ represents the total number of solutions that could have utilized that reflective check.}\label{tab:Creflection}
  \begin{ruledtabular}
    \begin{tabular} {l c c c}
	\textbf{Reflective check} 	& \textbf{N$_{total}$} & \textbf{N$_{incorrect}$} 	& \textbf{N$_{explicit}$}\\
	\hline
	 Units (R1)		& 125 		& 19	&  0		\\
           Boundary conditions (R2)       & 138		&  97	& 1  \\
           Laplace's equation	(R4) 	&  154			& 30	& 1  	\\
    \end{tabular}
  \end{ruledtabular}
\end{table}

Another strategy for understanding Reflection involves looking at the number of solutions where the final expression included an error that would have been detected by one or more of these checks.  Table \ref{tab:Creflection} lists this along with the number of solutions that explicitly included each reflective check.  Overall, these results suggest that an explicit check of boundary conditions would likely be the most effective reflective practice for students in terms of detecting errors, but that our students are rarely executing this (or other) checks spontaneously.

\subsection{\label{sec:resultsSoVs}\emph{Student difficulties with Separation of Variables in Spherical Coordinates}}

This section presents the identification and analysis of common student difficulties with the separation of variables technique in spherical coordinates organized by component and element of the operationalized ACER framework (Sec.\ \ref{sec:Sacer}).  

\subsubsection{\label{sec:activations}Activation of the tool}  

One instructor exclusively used A1-type prompts on both the midterm and final exams in his course (N=138); thus, these solutions provide minimal insight into Activation.  However, of the solutions to exam questions with implicit prompts (i.e., consistent with elements A2 or A3), very few (4\%, N=16 of 371) utilized a method other than SoV (e.g., Coulomb's law, $\vec{E}=-\vec{\nabla} V(R,\theta)$, etc.).  In contrast, on the multiple-response CUE question (see Sec.\ \ref{sec:context}) asked at the end of the semester (but before the final) just under half of our students (41\%, N=59 of 145) did not select SoV as the appropriate solution method.  This trend is slightly increased for other institutions with just over half the students selecting other methods (60\%, N=96 of 161).  The most common alternatives were Direct Integration via Coulomb's law (26\%, N=40 of 155 incorrect responses, all institutions) and Gauss' law (45\%, N=69 of 155 incorrect responses, all institutions).  This may be a reflection of the fact that spherical SoV questions, while still distinctive from an expert point of view, are potentially less recognizable to students than their Cartesian counterparts, and their superficial similarity to problems that might be solved by Coulomb's law or Gauss' law may discourage students from activating their SoV resources.  

The second set of interviews provided additional insight into Activation of spherical SoV through a question like the one shown in Fig.\ \ref{fig:examqSoVs}.  Of the six interview participants, three spontaneously brought up Laplace's equation and suggested SoV as the correct solution method.  Of the remaining students, one mentioned Laplace's equation only after being prompted to consider the fact that all the charges would be confined to the surface of the shell, while the other two needed to be explicitly told to consider Laplace's equation.  Moreover, these three students only suggested using SoV after being reminded that Laplace's equation is a non-trivial PDE and asked how we generally deal with PDEs in physics.  This result may suggest that, as we might expect, the activation of resources related to SoV for these students was more closely linked to the formal mathematics of the problem (i.e., solving a PDE), rather than the physical context (i.e., solving for the voltage).  

Students' overall success at Activation on the midterm and final exam questions seems to contradict the significantly lower success rate seen on the CUE and in interviews.   One potential explanation for this is that students may be simply pattern matching on the exams rather than internalizing a clear motivation for when and why SoV is appropriate.  The proximity of the exams to classroom instruction on SoV can make pattern matching a highly effective strategy.  This interpretation is supported by the following comment made by one of the interview participants: ``I remember these questions; I used to love these questions, and I don't remember how to do them anymore ... I guess I didn't understand this problem as well as I should have; I just remember going through a mathematical, like, process to get it, and I knew that one really well.''

\subsubsection{\label{sec:constructions}Construction of the model} 

The Construction component deals with mapping between the physics and mathematics of a problem.  For spherical SoV, this process includes identifying all necessary boundary conditions (element C1), both those provided explicitly in the prompt and those that are implicit in the underlying physics of localized charge distributions (i.e., $V(r\rightarrow \infty) \rightarrow 0$ and $V(r \rightarrow 0) \neq \infty$).  Of the solutions that utilized SoV on the exams (N=488), almost two-thirds (61\%, N=298 of 488) included correct expressions for all explicit and implicit boundary conditions.  Of the remaining solutions, more than half (62\%, N=117 of 190) never expressed the relevant implicit boundary conditions at $r=0$ and/or $r=\infty$.  Despite this, the majority of these solutions (89\%, N=104 of 117) correctly eliminated either the $A_l$ (outside) or $B_l$ (inside) terms.  This move was often accompanied by seemingly axiomatic statements like ``$A_l$'s go to zero outside.''  This finding is also consistent with the hypothesis that some students are using pattern matching to guide their solution rather than clearly justifying their steps from the underlying physics.  Other issues with expressing the boundary conditions (element C1) included using incorrect or inappropriate implicit boundary conditions (12\%, N=22 of 190, e.g., enforcing $V(r\rightarrow \infty) \rightarrow 0$ when solving for V inside a sphere), or incorrectly expressing the surface boundary condition (22\%, N=42 of 190, e.g., arguing $V(R)=V_o cos^2\theta \rightarrow V_o P_2(cos\hspace{1mm}\theta)$).  

As upper-division students are rarely (if ever) expected to derive the general expression for the voltage from Laplace's equation in spherical coordinates, the second element of Construction does not typically apply to spherical SoV questions.  Alternatively, the final element of Construction deals with setting up the equations to solve for the unknown constants in the general solution in order to match the boundary conditions (note, algebraic mistakes related to solving these equations will be discussed in relation to the Execution component).  Ultimately, just under a fifth of solutions (19\%, N=90 of 485) included issues with setting up the equations to solve for one or more constants.  The most common issues included (see Table \ref{tab:Ssc3}): not plugging in $r=R$ when matching the boundary condition at the surface, problems expressing or eliminating $P_l$ terms, and including both $A_l$'s and $B_l$'s when matching the boundary condition at the surface despite previously setting $A_l$ or $B_l$ to zero.   	

\begin{table}
\caption{Difficulties setting up expressions match the boundary conditions in spherical. Percentages are of just the students who had difficulties setting up the boundary conditions (19\%, N=90 of 485).  Codes are not exhaustive or exclusive but represent the most common themes, thus the total N in the table need not sum to 90.}\label{tab:Ssc3}
  \begin{ruledtabular}
    \begin{tabular} {p{6.5cm} c c}
	\textbf{Difficulty} 				& \textbf{N} 	& \textbf{Percent}		\\
	\hline
	 Not setting $r=R$ for the surface boundary	& 21 		& 23\%				\\
           Problems with $P_l$ terms	 	& 16 			& 18\%					\\
	\hspace{3mm} e.g., expressing $P_l$'s incorrectly, dropping $P_l$ 	 & 	& 	\\
           \hspace{10mm} terms inappropriately  &    & 							\\
	Including both $A_l$'s and $B_l$'s in one expression  	& 10 			& 11\%		\\
	\hspace{3mm} e.g., $(A_l R^l+\frac{B_l}{R^{l+1}})P_l=V_o P_l$ 	& 	& 	\\
           Never applied the surface boundary condition & 18 & 20\% 					\\
    \end{tabular}
  \end{ruledtabular}
\end{table}

In interviews, students tended to move quickly back and forth between identifying boundary conditions and setting up equations to match them.  For example, all five participants who solved the spherical SoV question began by identifying one of the two implicit boundary conditions (element C1) and using it to correctly eliminate either the $A_l$ or $B_l$ terms (element C3).  All of these students then moved on to matching the boundary condition at the surface without commenting on either the second implicit boundary condition or what region their expression would be valid for.  Three interviewees correctly set up an expression to match the surface boundary condition (element C3).  One of the remaining students did not plug in $r=R$ into his expression until prompted, while the other student did not initially isolate like terms when solving for constants.  When asked where their final expression was valid, all five interviewees initially argued it would be valid everywhere.  Once they were specifically directed to consider limiting values of $r$, all interviewees recognized their solution was inconsistent with the remaining implicit boundary conditions, but only one student spontaneously considered the possibility of having separate expressions for $V(r)$ inside and outside the sphere.  Thus, the interviews suggest the tendency of both exam and interview students to not spontaneously acknowledge some or all of the implicit boundary conditions may discourage them from recognizing that their solution is valid only for certain regions of space or \emph{vice versa}.  

\subsubsection{\label{sec:executions}Execution of the mathematics}

The Execution component deals with the procedural aspects of working through the mathematics of a physics problem.  The first two elements of Execution address the process of separating Laplace's equation into ODEs by assuming the separated form of the potential (i.e., $V(r,\theta)=R(r)\Theta(\theta)$).  In spherical coordinates, this process yields a single, general solution for the potential.  Students in junior electrostatics are typically shown this derivation once and are rarely (if ever) expected to replicate it.  Thus elements E1 and E2 are not typically necessary for problems involving spherical SoV.  

Once a student has used the boundary conditions to set up expressions for the unknown constants (element C3), there are any number of mathematical manipulations that may be necessary to solve for these constants (element E3).  As described previously, we have noted three common strategies that can be used in this process (see Sec.\ \ref{sec:resultsSoVc}): zero matching, Fourier's trick, and term matching.  Of the exam solutions that showed explicit evidence of Execution (92\%, N=469 of 509), nearly all (97\%, N=455 of 469) used some form of zero matching to eliminate one set of constants ($A_l$'s or $B_l$'s).  The majority of solutions also used term matching (89\%, N=405 of 455) to solve for the non-zero constants, while only a small fraction (12\%, N=56 of 455) used Fourier's trick.  This strong preference for term matching is appropriate and is likely a reflection of the fact that nearly all surface boundary conditions given on exams at CU can be expressed as a sum of 1-3 Legendre polynomials.  Moreover, several of the exam prompts provide or explicitly ask students to express the boundary condition, $V(R)$, in terms of Legendre polynomials.  

\begin{table}
\caption{Common difficulties when executing the procedural mathematics of solving for constants in the general solution. Percentages are of just the students who exhibited these difficulties (27\%, N=125 of 469).  Codes are not exhaustive or exclusive but represent the most common themes, thus the total N in the table need not sum to 125.}\label{tab:Sse3}
  \begin{ruledtabular}
    \begin{tabular} {p{6.6cm} c c}
	\textbf{Difficulty} 				& \textbf{N} 	& \textbf{Percent} 					\\
	\hline
           Incorrect term matching 	 	&  32			& 26\%		\\
	\hspace{3mm} e.g., keeping too many or not enough $P_l$'s	 & 	& 		\\
	 Off by a unitless constant factor or sign		& 27 		&  22\%			\\
           \hspace{3mm} e.g., factor of 2       & 		& 	\\
           Off by a unitfull constant factor or sign		& 22 		&  18\%			\\
           \hspace{3mm} e.g., $V_o$, or radius $R$       & 		& 	\\
	Not finishing the calculation  	& 24 			& 19\%				\\
    \end{tabular}
  \end{ruledtabular}
\end{table}

When solving for the values of the unknown constants (element E3), roughly a quarter of students' solutions (27\%, N=125 of 469) contained various mathematical mistakes.  Common issues included (see Table \ref{tab:Sse3}): losing or gaining a constant factor, keeping or losing $P_l$ terms inconsistent with the boundary condition, and not finishing the calculation.  The fraction of solutions with mathematical errors was higher in solutions that utilized Fourier's trick to determine non-zero constants (60\%, N=34 of 56) than in solutions that utilized term matching (21\%, N=84 of 405).  This is likely due, at least in part, to the fact that Fourier's trick represents an inherently more mathematically demanding strategy.  

The final element in Execution (element E4) deals with compiling all aspects of the solution into a single expression for the voltage.  Roughly three-quarters of the solutions (73\%, N=374 of 509) were completed enough to potentially include a final expression for the voltage, and most (83\%, N=313 of 374) did so correctly.  Common mistakes included not compiling a final expression (23\%, N=14 of 61), dropping or adding terms (25\%, N=15 of 61), and not including the $r$-dependence from the general solution (21\%, N=13 of 61).  Ultimately, only a small fraction of students (8\%, N=43 of 509) had difficulties \emph{only} with elements of the Execution component (i.e., no mistakes in Activation or Construction).

The interviews provided relatively minimal insight into student difficulties in the Execution component, in part because only two of the five students made mathematical errors of any kind while solving the spherical SoV problem.  Both of these students initially failed to include the $r$-dependence from the general solution when compiling their expression for the voltage.  Comments made by these two students suggested that they were focusing on how their final expression matched the boundary condition at $r=R$.  As the boundary condition does not have $r$-dependence, this may account for these students leaving the $r$-dependence out of their final expression.  The overall success of the interviewees with respect to Execution may be due in part to both the simplicity of the given boundary condition ($V(R)=V_o(1+cos\hspace{1mm}\theta)$) and the fact that all of the interviewees used term matching rather than Fourier's trick to solve for the non-zero constants.  Thus the mathematical manipulations required for this problem were minimal and purely algebraic.  Overall, analysis of both the interviews and exam solutions suggest that Execution rarely represents the primary barrier to student success on spherical SoV problems.  

\subsubsection{\label{sec:reflections}Reflection on the Result}

\begin{table}
\caption{Number of exam students who explicitly utilized each of the four possible reflective checks (N$_{explicit}$) along with the number of solutions that included an error that would have been detected by this check (N$_{incorrect}$).  N$_{total}$ represents the total number of solutions that reached a point where they could have utilized that reflective check.  The limiting behavior check applies only to exams that asked for $V_{outside}$, and the two semesters in which students were directly prompted to consider limiting behavior are excluded; this accounts for the lower N$_{total}$.  }\label{tab:Sreflection}
  \begin{ruledtabular}
    \begin{tabular} {l c c c}
	\textbf{Reflective check} 	& \textbf{N$_{total}$} 	 & \textbf{N$_{incorrect}$} & \textbf{N$_{explicit}$}\\
	\hline
	 Units (R1)		& 315 		& 39	&  2		\\
           Boundary conditions (R2)       & 382	  &  119	& 22	\\
           Limiting behavior (R3)  &  157 & 27  &  5  \\
           Laplace's equation (R4) 	&  365			& 51	& 1  	\\
    \end{tabular}
  \end{ruledtabular}
\end{table}

We identified four reflective checks that a student could use to gain confidence in (or detect problems with) their solution to problems involving spherical SoV (elements R1-4).  Only a small fraction of our students made explicit, spontaneous attempts to check their final expressions (7\%, N=27 of 397) and the majority of these did so \emph{only} by checking boundary conditions (70\%, N=19 of 27).  In interviews, two of five students made spontaneous attempts to check their solution, one by looking at units and the other at boundary conditions.  One additional student suggested checking units after being asked how he might convince himself his solution was correct.  

Two of the exam prompts directly targeted element R3 by asking students to comment on why they might expect the first term in the potential outside the sphere to behave as $1/r$ (the given surface voltage was everywhere positive).  A completely correct response requires that the student recognize that if the voltage is everywhere positive, then the sphere must have net positive charge on its surface, and thus would look like a point charge in the limit of large $r$.  Of the solutions to these two exams, only a small fraction (8\%, N=6 of 72) articulated this relatively subtle argument fully.  Common alternative justifications included that $1/r$ was the dependence for a point charge but made no comment about the charge on the sphere (46\%, N=33 of 72), or that $1/r$ goes to zero at infinity which matches the boundary condition (13\%, N=9 of 72).  Similarly, all three of the interview students who were asked about limiting behavior of the potential needed explicit guidance before recognizing that the overall sign of the potential can be used to infer the sign of the total charge.  If a significant fraction of our students struggled to produce an expectation for the behavior of the potential at large $r$, this likely contributed to why checks of limiting behavior were so rare.  

It is also possible that more of the exam students performed one of these reflective checks spontaneously, but did not explicitly write it down on their solution.  To address this, we can also examine the fraction of solutions that included errors in their final expressions that would have been detected by one or more of these checks.  Table \ref{tab:Sreflection} lists this along with the number of students who explicitly utilized each reflective check.  These results suggest that explicit checks of boundary conditions are both the most common (though still rare) and potentially the most effective in terms of catching errors.  

\subsection{\label{sec:implications}Implications for Instruction}

While it was not the goal of this study to investigate the impacts of different instructional strategies or curricular materials on the prevalence or persistence of students' difficulties with separation of variables, our findings do suggest several implications for teaching SoV in electrostatics.  First, for problems in Cartesian coordinates, both the introduction of the separated form of the potential and the infinite sum are critical pieces of the solution that students have difficulty clearly justifying and/or coming up with spontaneously.  It may be particularly important to directly target these two issues in order for students to form a more robust conception of the SoV technique.  For example, getting students to come up with the need for the infinite sum on their own (i.e., asking them to try solving for the final non-zero constant(s) without it), rather than simply telling them it was necessary, seemed to be a particularly productive exercise for our interview students.  

It is also worth acknowledging that solving SoV problems solely through pattern matching, while undesirable in terms of generalizability, is often a highly effective strategy, in part because there are a finite number of solvable SoV questions and they are all fairly similar.  However, we have identified several variations on these canonical questions that may help to discourage students from purely pattern matching, particularly for Cartesian questions.  For example, placing the non-zero boundary condition on either the $y=0$ or $x=0$ sides of a rectangular box can complicate the simplification of the exponential term.  Alternatively, providing a function rather than a constant for the non-zero boundary condition (i.e., $V(x,y=0)=V_osin(\pi x/a)$) can also force students to adapt their normal Cartesian SoV procedure.  This latter strategy would also provide an avenue for an explicit discussion of when Fourier's trick vs.\ term matching represent the most efficient strategy for solving for the unknown constants.  For spherical SoV, asking for the potential between two nested spherical shells may also discourage pattern matching as neither the $A_l$'s or $B_l$'s go to zero in this case.  

\section{\label{sec:discussion}Summary and Implications}

We investigated upper-division student difficulties when using the separation of variables technique to solve Laplace's equation in the context of junior electrostatics by examining students' solutions to exam questions, a conceptual post-test, and think-aloud student interviews.  We found that our students encountered a number of identifiable issues when solving SoV problems, and that these difficulties differed for problems involving spherical and Cartesian geometries.  The ACER framework helped us to organize and categorize these difficulties within the problem-solving process.  

For Cartesian SoV, we found that our students were highly successful in terms of recognizing SoV as the appropriate mathematical technique when presented with canonical Cartesian SoV questions.  Alternatively, a subset of our students used a general expression for the potential that did not satisfy Laplace's equation, possibly because they were over-focusing on satisfying the boundary conditions.  Moreover, despite a relatively high rate of success on exams, we observed a number of issues in interviews relating to recalling/justifying the separated form of the potential (i.e, $V(x,y)=X(x)Y(y)$), applying the correct logic to separate Laplace's equation into several ODEs, and recalling/justifying the need for the infinite sum.  We suspect that this apparent disconnect between student performance on exams and interviews may be a reflection of students pattern matching on exams without being able to fully justify their steps.  Pattern matching, as a less robust strategy, is unlikely to be as effective in the interviews which take place after the course was completed.  

We also found that when solving for non-zero constants in Cartesian SoV, our students had a strong preference for Fourier's trick over term matching when both strategies were possible.  This preference, while understandable given the types of boundary conditions students are accustomed to, may have exacerbated student difficulties with the procedural mathematics as Fourier's trick represents a more mathematically demanding strategy.  Finally, very few of our students made spontaneous attempts to reflect on their solutions despite the fact that these strategies, particularly checking boundary conditions, can be highly effective at detecting errors made earlier in the solution.  

For spherical SoV, we found that our students were sometimes less successful in terms of recognizing SoV as the appropriate mathematical technique than for Cartesian SoV.  Interviews suggest that this may be due, in part, to a failure to activate Laplace's equation as the underlying equation that needs to be solved.  Many of the issues that arose in Cartesian SoV related to working through the process of separating Laplace's equation and solving the resulting ODEs; however, as students are typically not required to work through this process in spherical, these difficulties were not observed for spherical SoV.  

We also found that many students did not recognize and/or spontaneously identify all implicit boundary conditions on the potential in spherical coordinates.  As boundary conditions are typically explicitly given in Cartesian SoV, this difficulty was not observed in Cartesian problems.  Moreover, we found that in contrast to the tendency to use Fourier's trick for Cartesian SoV, our students had an appropriate preference for term matching when solving for non-zero constants in spherical SoV problems.  Students' tendency to prefer one strategy over the other is likely a reflection of the canonical kinds of boundary conditions that are used for these two different geometries.  However, this tendency may also suggest that students have not explicitly connected term matching and Fourier's trick as related strategies, which may be a symptom of a more fundamental difficulty with understanding and generalizing the properties of orthogonal functions.  Consistent with this and the idea that Fourier's trick represents a more mathematically demanding strategy,  fewer students had difficulty with the procedural mathematics in spherical problems that Cartesian.  

The idea that some students were solving SoV problems primarily through pattern matching was also supported by student work around spherical SoV.  For example, we consistently found that students make unjustified simplifications on spherical exam problems (e.g., setting $A_l$'s to zero without explanation), and one interview student made explicit comments about recalling that there was an explicit procedure for solving these problems but was not able to reproduce it.  Consistent with the results for Cartesian, we again found that students rarely made spontaneous attempts to reflect on their final solutions when solving spherical SoV problems despite making a number of errors that would have been detected through one or more of these checks.  

Additional work is needed to identify student difficulties when utilizing SoV in other contexts such as quantum mechanics, and the ACER framework represents a useful tool for facilitating comparisons of student problem-solving across contexts and courses.  Such investigations could also provide a longitudinal perspective on the growth of student understanding over time, allowing researchers and instructors to focus their efforts on addressing those difficulties that are most common and most persistent throughout the physics curriculum.

\begin{acknowledgments}
Particular thanks to the PER@C group and Marcos D. Caballero for their help and feedback.
This work was funded by NSF-CCLI Grant DUE-1023028 and an NSF Graduate Research Fellowship under Grant No. DGE 1144083.
\end{acknowledgments}

\bibliography{master-refs-04-01-15}
\bibliographystyle{apsper}   

\end{document}